\documentclass[conference,twoside]{IEEEtran}

\usepackage{cite}
\usepackage{amsmath,amssymb,amsfonts}
\usepackage{bbm}
\usepackage{algorithmic}
\usepackage{multirow}
\usepackage{graphicx}
\usepackage{comment}
\usepackage{textcomp}
\usepackage{xcolor}
\def\BibTeX{{\rm B\kern-.05em{\sc i\kern-.025em b}\kern-.08em
    T\kern-.1667em\lower.7ex\hbox{E}\kern-.125emX}}
\usepackage{lineno,hyperref}
\usepackage{caption}
\usepackage{subcaption}
\DeclareMathOperator*{\argmin}{arg\,min}

\usepackage{filecontents}

\ifCLASSINFOpdf
\else
\fi
\begin{document}
%
\title{Generalizable Denoising of Microscopy Images using Generative Adversarial Networks and Contrastive Learning}
%
%
%

\author{Felix Fuentes-Hurtado, Jean-Baptiste Sibarita and Virgile Viasnoff
\thanks{F. Fuentes-Hurtado is with the KNODIS Research Group, Universidad Politécnica de Madrid, Madrid, Spain, and also with the Departamento de Sistemas Informaticos, Universidad Politécnica de Madrid, Madrid, Spain.\\ 0000-0002-4320-245X}
\thanks{J. Sibarita is with the Centre national de la recherche scientifique (CNRS), Interdisciplinary Institute for Neuroscience (IINS), University of Bordeaux, France \\
0000-0002-9920-7700.}
\thanks{V. Viasnoff is with CNRS@CREATE, Mechanobiology Institute, Department of Biological Sciences, Singapore, and also with the National University of Singapore, Singapore\\
0000-0003-3949-2244.}
}

\maketitle

\begin{abstract}
Microscopy images often suffer from high levels of noise, which can hinder further analysis and interpretation. Content-aware image restoration (CARE) methods have been proposed to address this issue, but they often require large amounts of training data and suffer from over-fitting. To overcome these challenges, we propose a novel framework for few-shot microscopy image denoising. Our approach combines a generative adversarial network (GAN) trained via \textit{contrastive learning} (CL) with two structure preserving loss terms – Structural Similarity Index and Total Variation loss – to further improve the quality of the denoised images using little data. We demonstrate the effectiveness of our method on three well-known microscopy imaging datasets, and show that we can drastically reduce the amount of training data while retaining the quality of the denoising, thus alleviating the burden of acquiring paired data and enabling few-shot learning. The proposed framework can be easily extended to other image restoration tasks and has the potential to significantly advance the field of microscopy image analysis.
\end{abstract}

\begin{IEEEkeywords}
denoising, generative methods, contrastive learning, few-shot learning, microscopy data
\end{IEEEkeywords}

%
\IEEEpeerreviewmaketitle

\section{Introduction}
\label{sec:intro}

\IEEEPARstart{I}{mage} denoising has become a popular and crucial topic in computer vision problems due to the inherent presence of various types of noise in acquired images, which can significantly reduce their visual quality. Thus, removing noise from images is essential in numerous computer vision and image processing tasks \cite{liu2020connecting, fan2019brief, gu2019brief, goyal2020image}.

One of the key applications where image denoising is vital is microscopy imaging. Advanced microscopy techniques have enabled the investigation of biological processes at sub-cellular resolution, yet they still suffer from high noise levels during the acquisition process, which make the obtained images difficult to analyze and process. Therefore, image denoising is an essential step for accurate image analysis and interpretation in microscopy imaging.

The removal of unwanted noise and distortions from acquired images, known as \textit{image denoising}, is an important and challenging task in the field of computer vision. Mathematically, the image degradation process can be described as $x=y+n$, where $x$ represents the degraded form of the original image $y$ with added noise $n$. Although Gaussian noise (AWGN) is often used to model image noise \cite{chen2018image, krull2020probabilistic}, in reality, noise in images can be caused by various mechanisms, such as low-light noise, shot noise, readout noise and non-structured noise. Therefore, the development of denoising methods that can generalize to different types of noise is crucial.

Recent research has shown that Convolutional Neural Networks (CNNs) can effectively construct powerful and efficient content-aware image restoration pipelines, both supervised \cite{weigert2017isotropic, weigert2018content, zhang2017beyond, zhang2019poisson} and self-supervised \cite{batson2019noise2self, krull2019noise2void, krull2020probabilistic, lehtinen2018noise2noise, laine2019high}. Supervised approaches require corresponding pairs of noisy and clean images to learn a mapping between the two quality levels. Conversely, self-supervised methods perform this task without paired data by making assumptions about the characteristics of noise and signals.

Each method has advantages and disadvantages. Supervised methods learn directly from corresponding pairs of clean and noisy images to map the noisy images to their clean counterparts. This enables them to remove structured noise (artefacts) in addition to pixel-noise, and find the best possible solution for denoising. However, unsupervised or self-supervised methods are not dependent on clean images and can work with noisy images alone. While they cannot guarantee the best possible solution, they are still able to effectively remove pixel-noise, making them a preferred option in situations where obtaining clean counterparts is challenging.

Moreover, despite the benefits of unsupervised methods, most of them do not surpass the performance of supervised methods. At the time of writing this work, only two unsupervised methods have achieved better results than supervised methods \cite{prakash2020fully, prakash2021removing}. Interestingly, both methods employ variational auto-encoders (VAEs) and average several predictions, which results in blurrier images compared to those obtained by a single prediction made by supervised methods. Moreover, the sampling process employed by these networks makes them slower than other methods. Thus, the development of a generative model based on VAEs that produces sharp images remains an open and important question that requires further attention.

In our work, we aim to address the limitations of VAE-based methods in image restoration by exploring the potential of Generative Adversarial Networks (GANs)\cite{goodfellow2020generative}. GANs have shown great promise in generating sharp and high-quality images. We propose to use this learning framework to develop an efficient and effective method for image denoising. The GAN architecture consists of two networks - the generative network and the discriminative network. The generative network purpose is to generate samples that are indistinguishable from real data, while the discriminative network is trained to distinguish between real and generated samples. By leveraging GANs, we aim to produce high-quality denoised images that are visually appealing and closer to the ground truth.

Formally, GANs are generative models that learn a mapping from a random noise vector $z$ to an output image $y$, $G:z\rightarrow y$ \cite{goodfellow2020generative}.

One particular use of GANs is the image-to-image task, which consists on transforming an image defined in a given domain to its counterpart in a different domain \cite{isola2017image}. This method is suitable for paired data and can provide high-resolution, high-quality transformations, therefore deeming it suitable for the denoising task previously introduced. For this work, however, we are not feeding our generator with a random noise vector, but with a noisy image that we want to denoise instead. This type of problem is typically addressed with \textit{conditional} GANs (cGANs), which learn a mapping from an observed image $x$ and a random noise vector $z$, to a ``transformed'' image $y$, $G:{x,z} \rightarrow y$.

In this work, we present a novel approach to few-shot microscopy image denoising combining both supervised and unsupervised learning techniques. Concretely, our method is based upon a conditional Generative Adversarial Network built upon the well-known U-Net architecture trained via contrastive learning \cite{khosla2020supervised}. Moreover, we add two loss terms to increase the quality of the denoised images, often outperforming the current state of the art in image denoising. Finally, we demonstrate that it is possible to significantly reduce the number of training samples while maintaining high performance. To the best of our knowledge, our work is the first to use contrastive learning for microscopy image restoration. 

Our contributions are two-fold: (1) we present a conditional GAN based framework for microscopy image denoising based upon the U-Net architecture trained via contrastive learning with total variation loss and structural similarity index loss that complement each other and produce the highest quality; and (2) we demonstrate the robustness of our method, showing that it can effectively operate with limited data, specifically $\sim$10\% of the initial data, while still retaining a significant portion of the initial performance or even surpassing it.

\section{Related work}
\label{sec:related_work}

\subsection{Classical methods.} In the past few decades, a variety of filtering approaches has been used to address the denoising problem. Together with classical methods such as bilateral filtering \cite{tomasi1998bilateral} or Wiener filtering \cite{ghael1997improved}, the most prominent ones are Non-Local Means \cite{buades2005non}, which uses information from similar regions in the image to compute estimates of the underlying clean pixel values, and BM3D\cite{dabov2007image}, that leverages a two stage non-local collaborative filtering. The reader can find a detailed discussion and comprehensive survey of classical denoising methods in \cite{milanfar2012tour}.

\subsection{Deep Learning (DL) based methods.} Recent years have proven deep learning methods more effective in the image denoising task. DL based methods work by directly learning a mapping from a noisy image to its clean counterpart. This task can be performed in two ways: supervised or unsupervised. As discussed in Section \ref{sec:intro}, each approach has its advantages and disadvantages.

Some of the best-known supervised methods include the work of \cite{zhang2017beyond}, where a Deep CNN is used to perform blind Gaussian denoising by learning the residual between the clean and the noisy images; or that of \cite{weigert2018content}, that implements a simple U-Net architecture to learn the mapping between clean and noisy images and establishes a new baseline in microscopy image denoising. GANs have also been recently used for denoising tasks with success. For example, \cite{su2018generative} implement a GAN as a tool to recover structural information from cryo-electron microscopy data; \cite{gu2021robust} that apply a $\beta$-GAN combining GANs and auto-encoders to achieve a robust estimate of certain distributional parameters under Huber contamination model with statistical optimality; or \cite{ouyang2020research}, which also use a GAN-based method with a modified discriminator that performs regression and helps to stabilize the training process. Denoising Optical Coherence Tomography is also a task where GANs have been widely used. For instance, \cite{huang2019simultaneous} leverage a generative adversarial network combined with a pixel loss and a content loss in order to simultaneously denoise and augment the resolution of OCT images. \cite{huang2020noise} implement a GAN model using a set of encoders and decoders to disentangle the content of OCT images from the speckle noise. \cite{fuentes2022mid3a} employ a conditional GAN combined with differentiable data augmentation and two structure preserving losses to denoise microscopy images with few data.

On the unsupervised side, most methods address pixel-noise removal exclusively. Some of them need to be trained for each input image separately, such as \cite{jo2021rethinking} or \cite{quan2020self2self}. Most methods, however, can be trained on a corpus of noisy images and used on new, never-seen images without retraining the model, such as \cite{batson2019noise2self, krull2019noise2void, krull2020probabilistic, laine2019high, xie2020noise2same}. Recently, the works of Broadus \textit{et al.} \cite{broaddus2020removing} and Prakash \textit{et al.}\cite{prakash2021removing} showed that it is possible to successfully remove structured noise to some extent. There are a variety of works using GANs in an unsupervised manner, such as \cite{chen2018image}, who employ the GAN learning framework to generate pairs of noisy and clean images and apply employ supervised denoising networks; or \cite{manakov2019noise}, who use a cycleGAN to learn a mapping between noisy and clean images without paired images, even though they still need clean and noisy images. Or the ``cycle-free'' cycleGAN defined by \cite{kwon2021cycle}, which uses an invertible generator to be able to remove the need for paired data. 

\section{Method}
\label{sec:method}

The proposed method is built upon a conditional Generative Adversarial Network (i.e. pix-2-pix) modified to include two additional structured loss functions (Structural Similarity Index and Total Variation losses) to help denoise the images and trained via contrastive learning, to help the model learn the characteristics of negative and positive examples even when dealing with few data.

\subsection{Pix-2-pix architecture} 
\label{ssec:pix2pix}

We base our method in the pix-2-pix architecture, a kind of conditional GAN designed for Image-to-Image tasks. This architecture consists of a generator built upon the basic ``U-Net'' \cite{ronneberger2015u} architecture and a discriminator implementing a ``PatchGAN'' architecture \cite{isola2017image}. The ``U-Net'' is an encoder-decoder architecture with skip connections between mirrored layers in the encoder and decoder stacks. The choice of ``PatchGAN'' by \cite{isola2017image} as the architecture for the discriminator is based on the well known fact that the L2 and L1 losses produce blurry results on image generation problems \cite{larsen2016autoencoding}. These losses generally capture low frequencies, but fail at the high frequencies. Thus, the ``PatchGAN'' architecture is engineered so that it only penalizes structure at the scale of patches, thus focusing in high-frequency structure and relying on an L1 term to force low-frequency correctness. Hence, the image is modeled as a Markov random field, in which pixels separated by more than a patch diameter are assumed independent. Therefore, the original pix-2-pix architecture combines the adversarial loss and an L1 term to produce sharper images.

Our choice of this architecture for our work is motivated by the fact that we wanted an as-simple-as-possible backbone. In this manner, we intend to show that a simple approach combined with the right loss and regularization terms can perform just as well (if not better) than more complicated approaches.

\subsection{Adversarial loss}
\label{ssec:adv_loss}

The adversarial term of the conditional GAN objective of the pix-2-pix architecture can be expressed as:

\begin{equation}
    \label{eq:gan_objective}
    \begin{split}
    \mathcal{L}_{GAN}(G,D) & = \mathbb{E}_{x,y}[\log D(x,y)] \\
    & + \mathbb{E}_{x,z}[\log (1-D(x,G(x,z)))]
    \end{split}
\end{equation}

where $G$ tries to minimize this objective against an adversarial $D$ that tries to maximize it. After training this model, the trained generator model would be $G*=\argmin_{G} \max_{D} \mathcal{L}_{cGAN}$. 

However, for the pix-2-pix architecture, the authors chose to remove the random noise input since they found that the network was ignoring it. Instead, they add some stochasticity by including dropout on several layers of the generator both on training and testing time, but succeed only to a certain level. However, this is not a critical point for our purpose, since we only need a plausible mapping between noisy and clean images. The original pix-2-pix equation then becomes

\begin{equation}
    \label{eq:pix2pix_objective}
    \begin{split}
    \mathcal{L}_{GAN}(G,D) & = \mathbb{E}_{x,y}[\log D(x,y)]
    \\ & + \mathbb{E}_{x}[\log (1-D(x,G(x)))].
    \end{split}
\end{equation}

\subsection{Pixel-level loss}
\label{ssec:l1_loss}

Pixel-level losses are widely used in image-to-image tasks to reduce the pixel-to-pixel difference between the initial and the generated images. In this case, we chose the L1-norm loss, which calculates the L1 norm of the difference between the denoised image $y$ and the original, noisy image $x$:

\begin{equation}
    \label{eq:l1_loss}
    \mathcal{L}_{L1}(x, y) = || y - x ||_1,
\end{equation}

where $x$ denotes the original, noisy image, and $y$ the synthesized, denoised counterpart.

\subsection{Structural Similarity loss}
\label{ssec:ssi_loss}

The previous L1 loss is not enough to produce sharp, denoised images. L1 loss assumes that pixels are independent of each other, whereas images are highly structured (i.e. ordering of the pixels carry important information about the content of an image). By making this assumption, it is possible to obtain the same L1 loss irrespective of the correlation between the original image and the generated one, even though this correlation can have a strong impact on perceptual similarity \cite{wang2009mean}.

To solve this issue, we chose to add a term based on the Structural Similarity (SSIM) index \cite{wang2004image} to our loss function. Contrary to L1 loss, the SSIM index provides a measure of the similarity by comparing two images based on luminance, contrast and structural similarity information.

Let define $l(x,y)=\frac{2\mu_x\mu_y+C_1}{\mu_x^2+\mu_y^2 + C_1}$ as the luminance similarity, where $\mu_x=\frac{1}{N}\sum^N_{i=1}x_i$ and $C_1$ is a constant; $c(x,y)=\frac{2\sigma_x\sigma_y+C_2}{\sigma_x^2+\sigma_y^2 + C_2}$ as the constrast similarity, where $\sigma_x=(\frac{1}{N-1}\sum^N_{i=1}(x_i - \mu_x)^2)^\frac{1}{2}$ and $C_2$ is a constant; and $s(x, y)=\frac{\sigma_{xy} + C_3}{\sigma_x\sigma_y + C_3}$ as the structural information, where $\sigma_x=\frac{1}{N-1}\sum^N_{i=1}(x_i - \mu_x)(y_i - \mu_y)$ and $C_3$ is a constant. It is possible then to define the SSIM index as $SSIM(x,y)=l(x,y)\cdot c(x,y) \cdot s(x,y)$.

For convenience, we used the Structural dissimilarity for our optimization objective as defined in \cite{loza2006structural}:

\begin{equation}
    \label{eq:ssim_loss}
    \mathcal{L}_{SSIM}(x, y) = \frac{1-SSIM(x,y)}{2}.
\end{equation}

where the resulting value can vary between 0 (both images are the same) and 1 (images are very different).

\subsection{Total Variation Loss}
\label{ssec:tv_loss}

We already defined losses to take care of the structure of the image, but our problem is that these losses might neglect the existance of noise. For this purpose, we introduce a new loss term with the aim of reducing the variation of neighbouring pixels in the denoised image, which would most likely be caused by the presence of noise. 

The Total variation (TV) measure has been used for denoising during decades \cite{rudin1992nonlinear}. In the continuous domain, for a 1D function, TV computes an integral over the difference among neighboring values, leading to smoother outputs. Formally, the total variation of a differentiable function $f$, defined on an interval $[a,b] \subset \mathbb{R}$ is defined as $V^b_a=\int^b_a ||f'(x)dx$ if $f'$ is Riemann-integrable. For 1D discrete signals ($y=[Y_1, ..., y_N]$) can be defined as $TV(y)=\sum^{N-1}_{n=1} |y_{n+1}-y_n|$. 

For our purposes, since we are dealing with images, we employ the 2D TV as a term in out optimization objective:

\begin{equation}
    \label{eq:tv_loss}
    \mathcal{L}_{TV}(y) = \sum_{i,j}|y_{i+1,j}-y_{i,j}| + |y_{i,j+1}-y_{i,j}|.
\end{equation}

To simplify our optimization objective, we use the anisotropic version of TV, defined as the sum of horizontal and vertical gradients at each pixel.

\subsection{Contrastive Learning} 
\label{ssec:cl}

Contrastive learning \cite{chen2020simple} aims to learn a representation function $f(x)$ that maps input images $x$ to a latent feature space $\mathbb{R}^d$. Given a pair of augmented images $x_1$ and $x_2$, contrastive learning aims to bring their corresponding feature vectors $f(x_1)$ and $f(x_2)$ closer if they share the same identity, while pushing them further apart otherwise. This is achieved by minimizing a contrastive loss function that encourages the distance between the feature vectors to be small for positive pairs (images with the same identity) and large for negative pairs (images with different identities). 

Formally, let $x$ be an input image and $f(x)$ be its corresponding feature vector in $\mathbb{R}^d$. We can define the contrastive loss for a positive pair of examples $(i, j)$ as:

$$\ell_{i,j} = -\log\frac{\exp({sim(f(x_1), f(x_2))/\tau})}{\sum_{j=1}^{N}\mathbbm{1}_{[j\neq i]} \exp({sim(f(x_i),f(x_j))/\tau})}$$

where $sim(f(x_1), f(x_2))$ denotes the similarity between the feature vectors of $x_1$ and $x_2$, $\mathbbm{1}_{[j\neq i]}$ is an indicator function that is 1 if $j \neq i$ and 0 otherwise, $\tau$ is a temperature parameter that controls the smoothness of the distribution, and $N$ is the batch size, formed by $N/2$ examples and their augmented versions. The contrastive loss encourages the model to learn representations that group similar noisy images together and separate dissimilar noisy images apart. In the context of image denoising, this leads to an embedding space where similar noisy images (e.g., with the same type of noise) are close to each other, while dissimilar noisy images (e.g., with different types of noise) are far apart. By training a denoiser on top of the learned embeddings, the model can effectively denoise images even in scenarios where only a small amount of data is available.

\subsection{Full optimization objective}
\label{ssec:full_obj}

The full objective function of our denoising framework is finally defined as a weighted sum of all the losses from (\ref{eq:pix2pix_objective}) to (\ref{eq:tv_loss}):

\begin{equation}
    \label{eq:full_obj}
    \mathcal{L}=\lambda_{GAN}\mathcal{L}_{GAN} + \lambda_{L1}\mathcal{L}_{L1} + \lambda_{SSIM}\mathcal{L}_{SSIM} + \lambda_{TV}\mathcal{L}_{TV}
\end{equation}

where the weights of each loss was empirically set to balance their importance.

\section{Data, Experiments and Results}

\subsection{Implementation details}
\label{ssec:implementation}

We define a ``Unet-256'' consisting of 7 downsampling and 7 upsampling blocks. 
The input size is 256x256 pixels. The batch size for all the experiments is 32, except for the generalization experiments, when different number of batch size are used as specified. The optimizer used is Adam with learning rate set to $2e^{-4}$, and $\beta_1=0.5$ and $\beta_2=0.999$. The temperature parameter for contrastive learning is set to $\tau=0.1$. We trained the model for a total of 1000 epochs, with a linear decay scheduler for the last 500 epochs. Finally, the coefficients of the different loss terms (when used) are $\lambda_{GAN}=1$, $\lambda_{L1}=1$, $\lambda_{SSIM}=10$ and $\lambda_{TV}=1e^{-4}$. This combination of coefficients achieves the overall best performance. The code is publicly available at \href{https://github.com/ffuhu/clidim-microscopy-image-denoising}{https://github.com/ffuhu/clidim-microscopy-image-denoising}.

\subsection{Datasets and evaluation metrics}
\label{ssec:data}

\subsubsection{Datasets.} We employed the datasets acquired by Prakash \textit{et al.} \cite{prakash2020fully} publicly available. There are three datasets: $(i)$ \textit{Convallaria} data, consisting of 100 noisy images of the same \textit{Convallaria} section; $(ii)$ mouse \textit{skull nuclei} dataset, consisting of 200 noisy acquisitions of the same static mouse skull nuclei; and $(iii)$ mouse \textit{actin} data, consisting of 100 noisy realizations of the same static actin sample. In order to obtain the corresponding ground truth (i.e. clean counter-parts) we simply average all the samples.

\subsubsection{Evaluation metrics.} We employ three different measures in order to assess the quality of the denoising process: $(i)$ the Peak Signal-to-Noise Ratio (PSNR), $(ii)$ the Structural Similarity Index (SSIM) and $(iii)$ the Normalised Root Mean Squared Error (NRMSE).

\subsection{Experiments}
\label{ssec:experiments}

We design a set of experiments to test the effect of each feature included in our proposed method. The first experiment performed is considered as our method baseline, since we simply use the pix-2-pix architecture to denoise the images without any improvement. We then experiment adding TV loss, SSIM loss, contrastive learning, and the combination of all of them. 
The ablation study carried out show that training our model via contrastive learning is the adition that produces the highest increase in quality and helps with the generalization of the method. Finally, the TV and SSIM losses help to further increase the quality of the denoised images. Table \ref{tab:results_ns} shows the results for each dataset.

\begin{table*}[!h]
\centering
\resizebox{0.9\textwidth}{!}{%
\begin{tabular}{lllllllllll}
\hline
 &  & \multicolumn{3}{l}{\textit{\textbf{Convallaria}}} & \multicolumn{3}{l}{\textbf{Mouse \textit{actin}}} & \multicolumn{3}{l}{\textbf{Mouse \textit{skull nuclei}}} \\
\textbf{ns} & \textbf{experiment} & \textbf{PSNR} & \textbf{SSIM} & \textbf{NRMSE} & \textbf{PSNR} & \textbf{SSIM} & \textbf{NRMSE} & \textbf{PSNR} & \textbf{SSIM} & \textbf{NRMSE} \\ \hline
\multirow{5}{*}{\textbf{all}} & \textbf{baseline} & 35.59 & .9409 & .0023 & 32.11 & .8474 & .0015 & 40.70 & .9427 & .0057 \\
 & \textbf{TV} & 35.59 & .9390 & .0019 & 32.75 & .8583 & .0013 & 40.65 & .9425 & .0028 \\
 & \textbf{SSIM} & 35.70 & .9393 & .0014 & 32.76 & .8566 & .0009 & 40.71 & .9458 & .0049 \\ 
 & \textbf{CL} & 36.95 & .9556 & .0012 & \textbf{33.46} & \textbf{.8708} & \textbf{.0008} & 40.97 & .9521 & .0023 \\
 & \textbf{CL + TV + SSIM} & \underline{\textbf{37.04}} & \underline{\textbf{.9567}} & \underline{\textbf{.0010}} & 33.42 & .8734 & .0009 & \underline{\textbf{41.04}} & \underline{\textbf{.9532}} & \underline{\textbf{.0058}} \\ \hline
\multirow{3}{*}{\textbf{32}} & \textbf{baseline} & 35.69 & .9420 & .0023 & 32.55 & .8454 & .0009 & 40.81 & .9476 & .0030 \\
 & \textbf{CL} & 36.61 & .9531 & .0018 & 33.45 & .8718 & .0009 & 41.01 & .9528 & .0046 \\
 & \textbf{CL + TV + SSIM} & \textbf{36.92} & \textbf{.9564} & \textbf{.0016} & \textbf{33.45} & \textbf{.8733} & \textbf{.0007} & \textbf{41.02} & \textbf{.9533} & \textbf{.0057} \\ \hline
\multirow{3}{*}{\textbf{16}} & \textbf{baseline} & 36.60 & .9527 & .0014 & 32.54 & .8453 & .0007 & 40.69 & .9447 & .0017 \\
 & \textbf{CL} & 36.77 & .9554 & .0018 & 33.41 & .8710 & .0011 & 41.00 & .9526 & .0065 \\
 & \textbf{CL + TV + SSIM} & \textbf{36.96} & \textbf{.9566} & \textbf{.0013} & \underline{\textbf{33.49}} & \underline{\textbf{.8744}} & \underline{\textbf{.0010}} & 41.01 & .9529 & .0048 \\ \hline
\end{tabular}%
}
\caption{\textbf{Quantitative denoising results for our ablation test}. We compare the contribution of each loss to the performance of the model and how the amount of training data affects. Best values are highlighted in \textbf{bold} for experiments done with each amount of data (all, 32 and 16 samples). Best overall values are \underline{underlined}.}
\label{tab:results_ns}
\end{table*}

\begin{figure*}
     \centering 
     \begin{subfigure}[b]{0.19\textwidth}
         \centering
         \includegraphics[width=\textwidth]{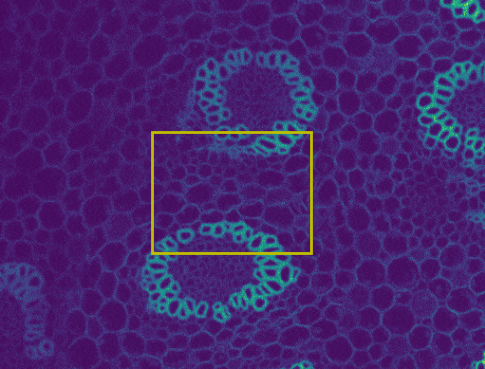}
     \end{subfigure}
     \begin{subfigure}[b]{0.19\textwidth}
         \centering
         \includegraphics[width=\textwidth]{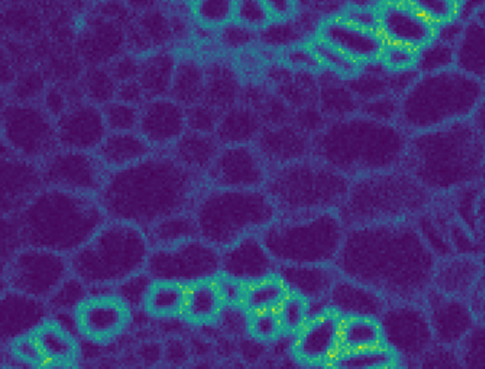}
     \end{subfigure}
     \begin{subfigure}[b]{0.19\textwidth}
         \centering
         \includegraphics[width=\textwidth]{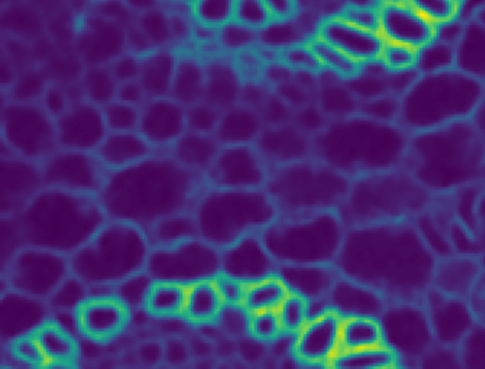}
     \end{subfigure}
     \begin{subfigure}[b]{0.19\textwidth}
         \centering
         \includegraphics[width=\textwidth]{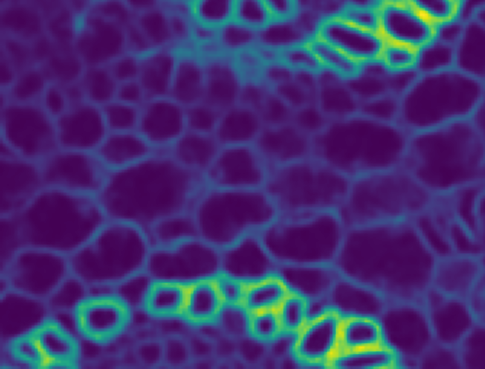}
     \end{subfigure}
     \begin{subfigure}[b]{0.19\textwidth}
         \centering
         \includegraphics[width=\textwidth]{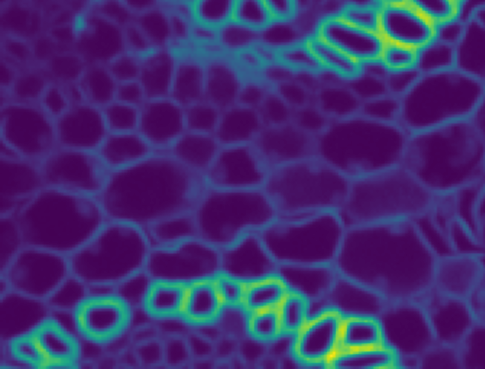}
     \end{subfigure}
     
     \par \smallskip
     
     \centering 
     \begin{subfigure}[b]{0.19\textwidth}
         \centering
         \includegraphics[width=\textwidth]{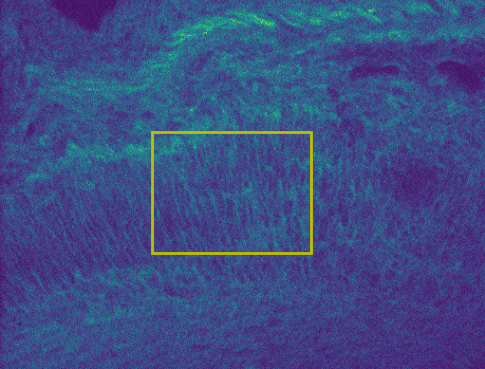}
     \end{subfigure}
     \begin{subfigure}[b]{0.19\textwidth}
         \centering
         \includegraphics[width=\textwidth]{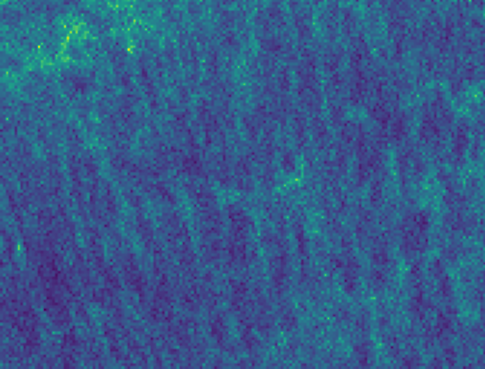}
     \end{subfigure}
     \begin{subfigure}[b]{0.19\textwidth}
         \centering
         \includegraphics[width=\textwidth]{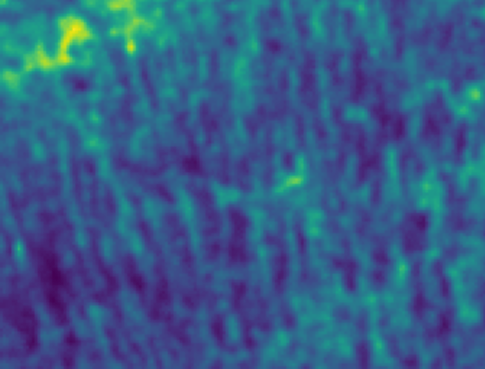}
     \end{subfigure}
     \begin{subfigure}[b]{0.19\textwidth}
         \centering
         \includegraphics[width=\textwidth]{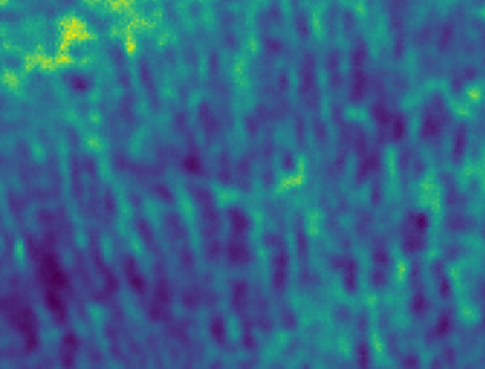}
     \end{subfigure}
     \begin{subfigure}[b]{0.19\textwidth}
         \centering
         \includegraphics[width=\textwidth]{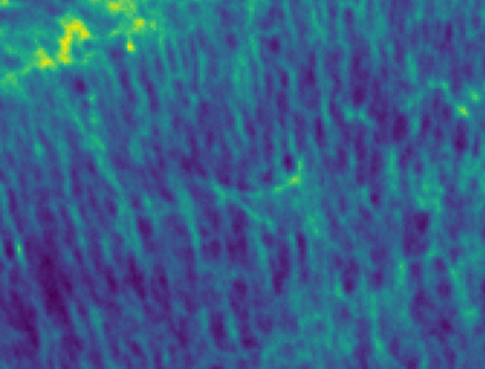}
     \end{subfigure}
     
     \par \smallskip
     
     \centering 
     \begin{subfigure}[b]{0.19\textwidth}
         \centering
         \includegraphics[width=\textwidth]{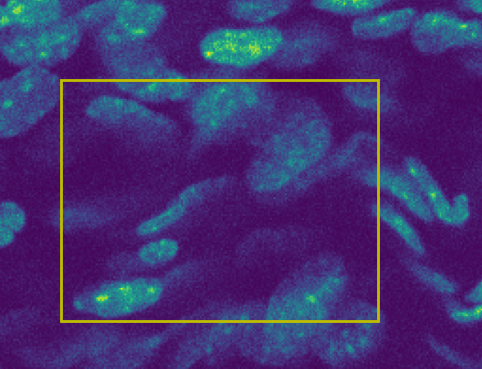}
        \caption{Noisy input}
     \end{subfigure}
     \begin{subfigure}[b]{0.19\textwidth}
         \centering
         \includegraphics[width=\textwidth]{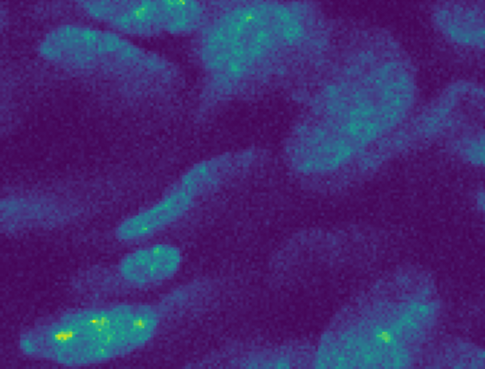}
         \caption{Input crop}
     \end{subfigure}
     \begin{subfigure}[b]{0.19\textwidth}
         \centering
         \includegraphics[width=\textwidth]{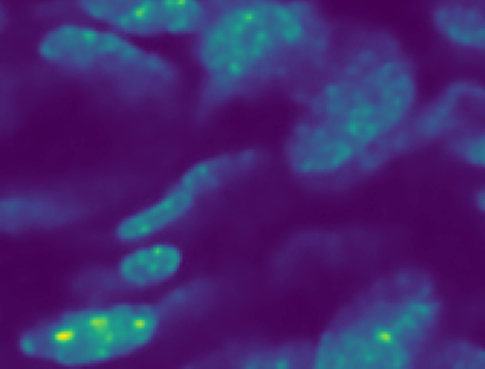}
        \caption{CARE}
     \end{subfigure}
     \begin{subfigure}[b]{0.19\textwidth}
         \centering
         \includegraphics[width=\textwidth]{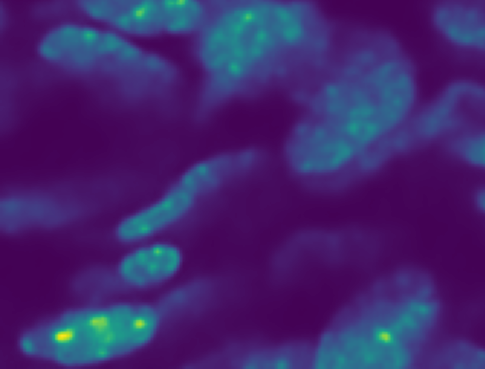}
        \caption{Our method}
     \end{subfigure}
     \begin{subfigure}[b]{0.19\textwidth}
         \centering
         \includegraphics[width=\textwidth]{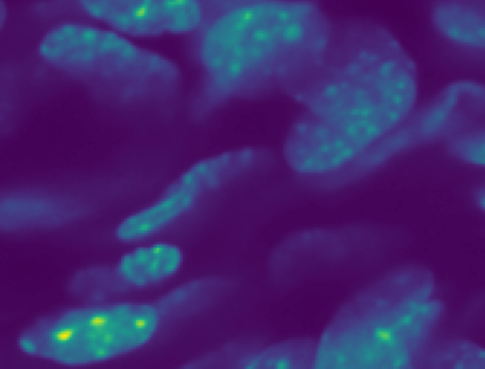}
        \caption{Ground truth}
     \end{subfigure}
     
    \caption{\textbf{Qualitative denoising results.} We compare the performance of our method against the current state-of-the-art method (CARE). Best viewed on a computer display.}
    \label{fig:qualitative_results}
\end{figure*}

We also include some samples to qualitatively evaluate the performance of the method (Figure \ref{fig:qualitative_results}).

Table \ref{tab:results_sota} shows our proposed method's performance in terms of PSNR compared to some of the current SOTA methods.
\begin{table*}[!h]
\centering
\resizebox{0.8\textwidth}{!}{%
\begin{tabular}{lcccccccc}
\hline
 & \multicolumn{4}{l}{\textbf{Unsupervised}} & \multicolumn{3}{l}{\textbf{Supervised}} \\
\textbf{Dataset} & \textbf{N2V} & \textbf{Vanilla VAE} & \textbf{DivNoising} & \textbf{PN2V} & \textbf{CARE} & \textbf{MID3A} & \textbf{Our method} \\ \hline
\textbf{\textit{Convallaria}} & 35.73 & 36.57 & 36.78 & 36.70 & 36.71 & \textbf{37.07} & 37.04 \\
\textbf{Mouse \textit{actin}} & 33.39 & 33.46 & 33.82 & 33.86 & \textbf{34.20} & 33.45 & 33.46 \\
\textbf{Mouse \textit{skull nuclei}} & 35.84 & 35.84 & 36.05 & 36.35 & 36.58 & 37.04 & \textbf{41.04} \\ \hline
\end{tabular}%
}
\caption{\textbf{Quantitative comparison with state-of-the-art methods in terms of PSNR measure.} Best values are highlighted in bold for each configuration and dataset.}
\label{tab:results_sota}
\end{table*}

\begin{figure*}
     \centering 
     \begin{subfigure}[b]{0.19\textwidth}
         \centering
         \includegraphics[width=\textwidth]{images/Convallaria_diaphragm_baseline_seed0_0_0_real_A_crop_200_328_160_328.png}
     \end{subfigure}
     \begin{subfigure}[b]{0.19\textwidth}
         \centering
         \includegraphics[width=\textwidth]{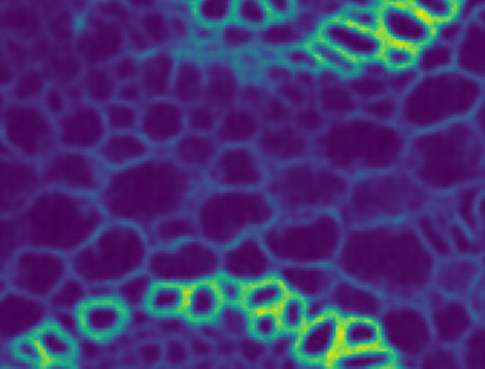}
     \end{subfigure}
     \begin{subfigure}[b]{0.19\textwidth}
         \centering
         \includegraphics[width=\textwidth]{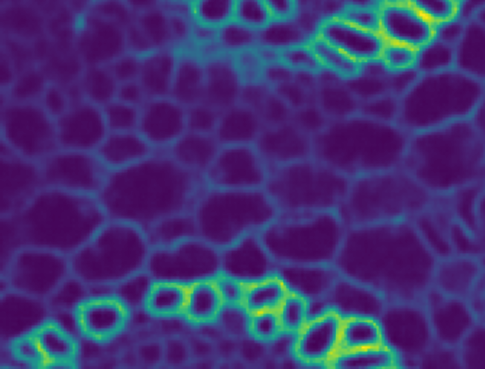}
     \end{subfigure}
     \begin{subfigure}[b]{0.19\textwidth}
         \centering
         \includegraphics[width=\textwidth]{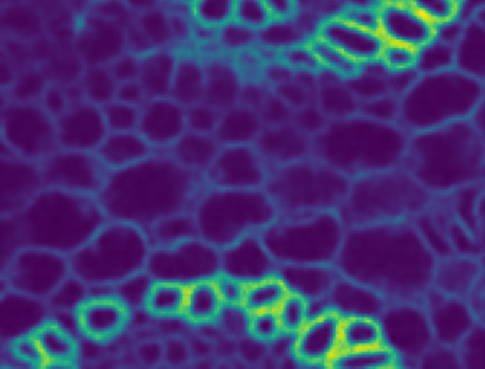}
     \end{subfigure}
     \begin{subfigure}[b]{0.19\textwidth}
         \centering
         \includegraphics[width=\textwidth]{images/Convallaria_diaphragm_baseline_seed0_0_0_real_B_crop_200_328_160_328.png}
     \end{subfigure}
     
     \par \smallskip
     
     \centering 
     \begin{subfigure}[b]{0.19\textwidth}
         \centering
         \includegraphics[width=\textwidth]{images/Convallaria_diaphragm_baseline_seed0_0_0_real_A_crop_200_328_160_328.png}
     \end{subfigure}
     \begin{subfigure}[b]{0.19\textwidth}
         \centering
         \includegraphics[width=\textwidth]{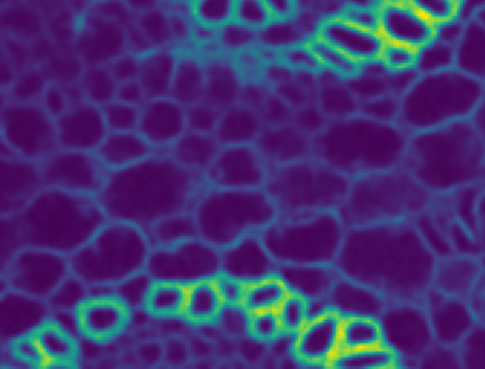}
     \end{subfigure}
     \begin{subfigure}[b]{0.19\textwidth}
         \centering
         \includegraphics[width=\textwidth]{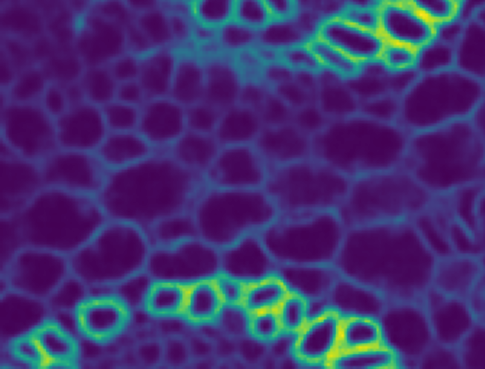}
     \end{subfigure}
     \begin{subfigure}[b]{0.19\textwidth}
         \centering
         \includegraphics[width=\textwidth]{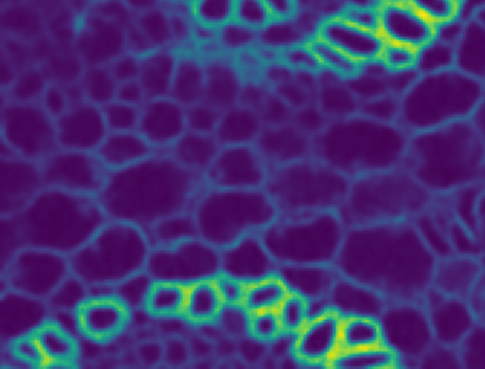}
     \end{subfigure}
     \begin{subfigure}[b]{0.19\textwidth}
         \centering
         \includegraphics[width=\textwidth]{images/Convallaria_diaphragm_baseline_seed0_0_0_real_B_crop_200_328_160_328.png}
     \end{subfigure}
     
     \par \smallskip
     
     \centering 
     \begin{subfigure}[b]{0.19\textwidth}
         \centering
         \includegraphics[width=\textwidth]{images/Convallaria_diaphragm_baseline_seed0_0_0_real_A_crop_200_328_160_328.png}
        \caption{Input crop}
     \end{subfigure}
     \begin{subfigure}[b]{0.19\textwidth}
         \centering
         \includegraphics[width=\textwidth]{images/Convallaria_diaphragm_CLreg0_TV1e-4_SSIM10_seed0_0_0_fake_B_crop_200_328_160_328.png}
         \caption{All training samples}
     \end{subfigure}
     \begin{subfigure}[b]{0.19\textwidth}
         \centering
         \includegraphics[width=\textwidth]{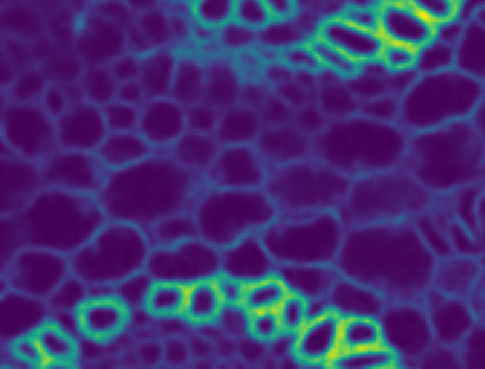}
        \caption{32 training samples}
     \end{subfigure}
     \begin{subfigure}[b]{0.19\textwidth}
         \centering
         \includegraphics[width=\textwidth]{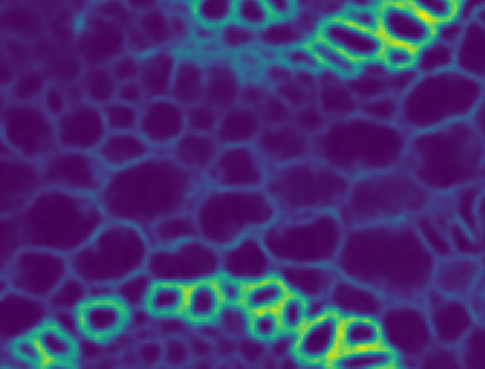}
        \caption{16 training samples}
     \end{subfigure}
     \begin{subfigure}[b]{0.19\textwidth}
         \centering
         \includegraphics[width=\textwidth]{images/Convallaria_diaphragm_baseline_seed0_0_0_real_B_crop_200_328_160_328.png}
        \caption{Ground truth}
     \end{subfigure}
     
    \caption{\textbf{Generalization study for the \textit{Convallaria} dataset}. First row shows the baseline configuration of our method, second row the version with contrastive learning, and the third row the complete proposed method. Best viewed on a computer display.}
    \label{fig:qualitative_results_ns_conv}
\end{figure*}

\begin{figure*}
     \centering 
     \begin{subfigure}[b]{0.19\textwidth}
         \centering
         \includegraphics[width=\textwidth]{images/Mouse_actin_baseline_seed0_0_0_real_A_crop_200_328_160_328.png}
     \end{subfigure}
     \begin{subfigure}[b]{0.19\textwidth}
         \centering
         \includegraphics[width=\textwidth]{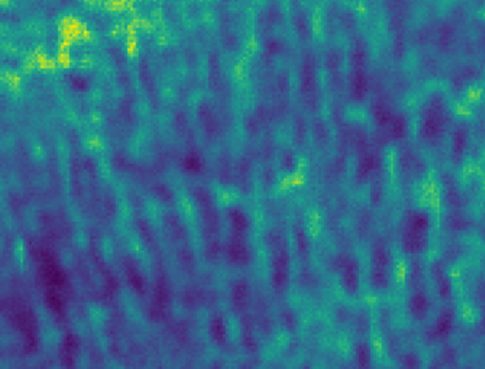}
     \end{subfigure}
     \begin{subfigure}[b]{0.19\textwidth}
         \centering
         \includegraphics[width=\textwidth]{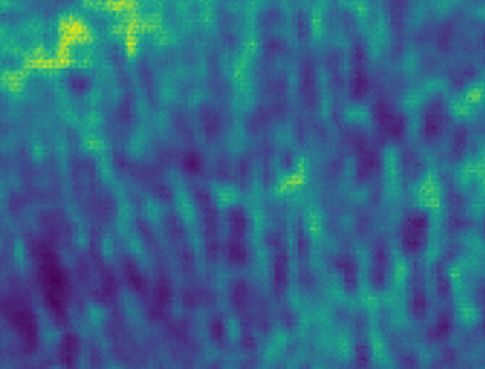}
     \end{subfigure}
     \begin{subfigure}[b]{0.19\textwidth}
         \centering
         \includegraphics[width=\textwidth]{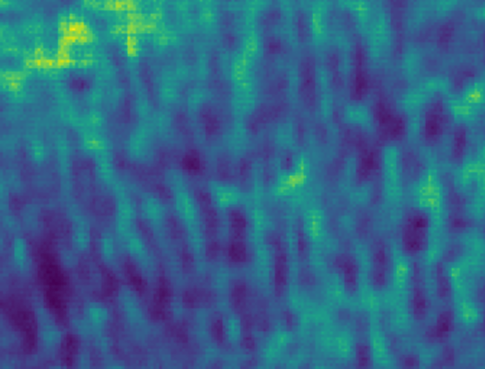}
     \end{subfigure}
     \begin{subfigure}[b]{0.19\textwidth}
         \centering
         \includegraphics[width=\textwidth]{images/Mouse_actin_baseline_seed0_0_0_real_B_crop_200_328_160_328.png}
     \end{subfigure}
     
     \par \smallskip
     
     \centering 
     \begin{subfigure}[b]{0.19\textwidth}
         \centering
         \includegraphics[width=\textwidth]{images/Mouse_actin_baseline_seed0_0_0_real_A_crop_200_328_160_328.png}
     \end{subfigure}
     \begin{subfigure}[b]{0.19\textwidth}
         \centering
         \includegraphics[width=\textwidth]{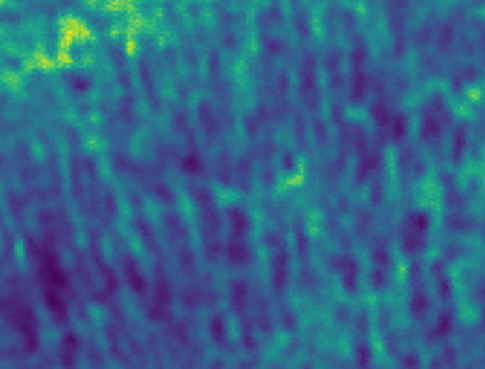}
     \end{subfigure}
     \begin{subfigure}[b]{0.19\textwidth}
         \centering
         \includegraphics[width=\textwidth]{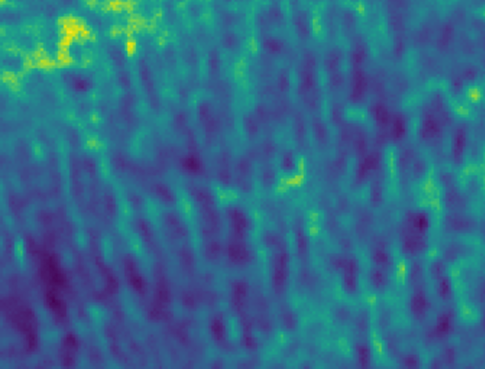}
     \end{subfigure}
     \begin{subfigure}[b]{0.19\textwidth}
         \centering
         \includegraphics[width=\textwidth]{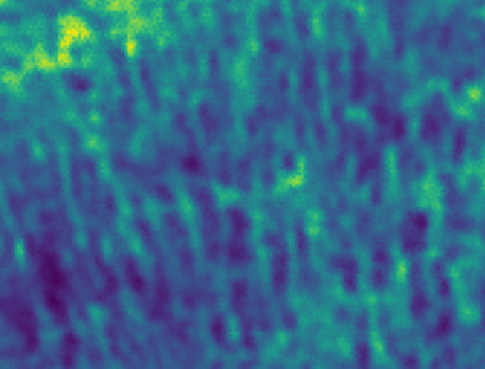}
     \end{subfigure}
     \begin{subfigure}[b]{0.19\textwidth}
         \centering
         \includegraphics[width=\textwidth]{images/Mouse_actin_baseline_seed0_0_0_real_B_crop_200_328_160_328.png}
     \end{subfigure}
     
     \par \smallskip
     
     \centering 
     \begin{subfigure}[b]{0.19\textwidth}
         \centering
         \includegraphics[width=\textwidth]{images/Mouse_actin_baseline_seed0_0_0_real_A_crop_200_328_160_328.png}
        \caption{Input crop}
     \end{subfigure}
     \begin{subfigure}[b]{0.19\textwidth}
         \centering
         \includegraphics[width=\textwidth]{images/Mouse_actin_CLreg0_TV1e-4_SSIM10_seed0_0_0_fake_B_crop_200_328_160_328.png}
         \caption{All training samples}
     \end{subfigure}
     \begin{subfigure}[b]{0.19\textwidth}
         \centering
         \includegraphics[width=\textwidth]{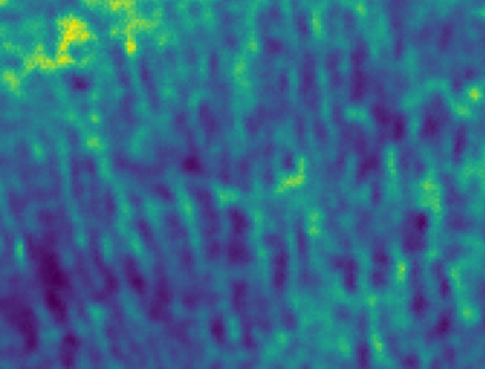}
        \caption{32 training samples}
     \end{subfigure}
     \begin{subfigure}[b]{0.19\textwidth}
         \centering
         \includegraphics[width=\textwidth]{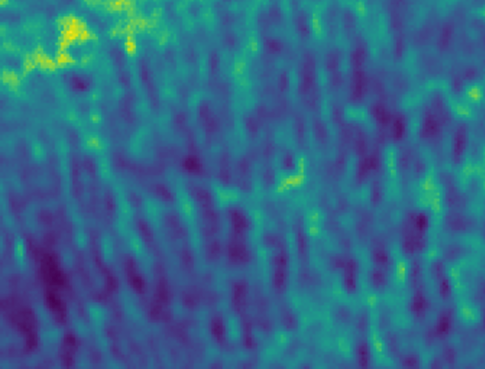}
        \caption{16 training samples}
     \end{subfigure}
     \begin{subfigure}[b]{0.19\textwidth}
         \centering
         \includegraphics[width=\textwidth]{images/Mouse_actin_baseline_seed0_0_0_real_B_crop_200_328_160_328.png}
        \caption{Ground truth}
     \end{subfigure}
     
    \caption{\textbf{Generalization study for the mouse \textit{actin} dataset}. First row shows the baseline configuration of our method, second row the version with contrastive learning, and the third row the complete proposed method. Best viewed on a computer display.}
    \label{fig:qualitative_results_ns_ma}
\end{figure*}

\begin{figure*}
     \centering 
     \begin{subfigure}[b]{0.19\textwidth}
         \centering
         \includegraphics[width=\textwidth]{images/Mouse_skull_nuclei_baseline_seed0_0_0_real_A_crop_72_200_32_200.png}
     \end{subfigure}
     \begin{subfigure}[b]{0.19\textwidth}
         \centering
         \includegraphics[width=\textwidth]{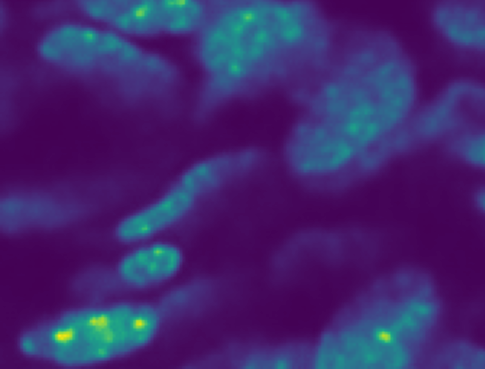}
     \end{subfigure}
     \begin{subfigure}[b]{0.19\textwidth}
         \centering
         \includegraphics[width=\textwidth]{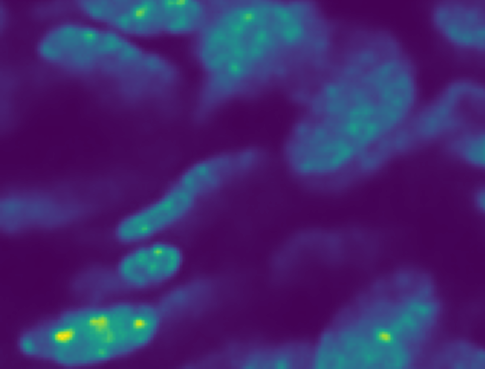}
     \end{subfigure}
     \begin{subfigure}[b]{0.19\textwidth}
         \centering
         \includegraphics[width=\textwidth]{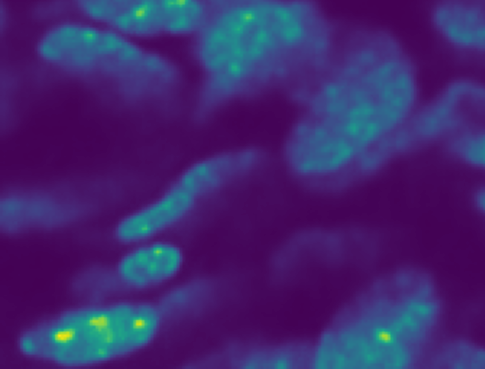}
     \end{subfigure}
     \begin{subfigure}[b]{0.19\textwidth}
         \centering
         \includegraphics[width=\textwidth]{images/Mouse_skull_nuclei_baseline_seed0_0_0_real_B_crop_72_200_32_200.png}
     \end{subfigure}
     
     \par \smallskip
     
     \centering 
     \begin{subfigure}[b]{0.19\textwidth}
         \centering
         \includegraphics[width=\textwidth]{images/Mouse_skull_nuclei_baseline_seed0_0_0_real_A_crop_72_200_32_200.png}
     \end{subfigure}
     \begin{subfigure}[b]{0.19\textwidth}
         \centering
         \includegraphics[width=\textwidth]{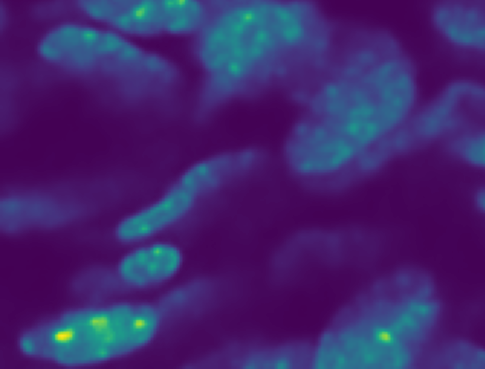}
     \end{subfigure}
     \begin{subfigure}[b]{0.19\textwidth}
         \centering
         \includegraphics[width=\textwidth]{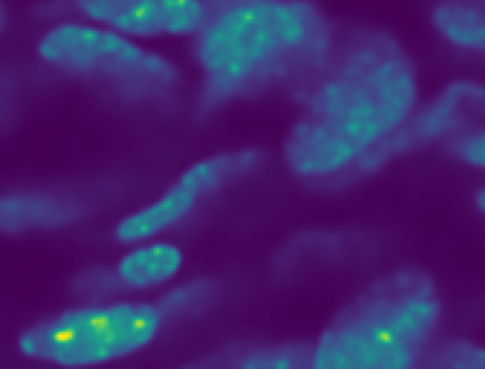}
     \end{subfigure}
     \begin{subfigure}[b]{0.19\textwidth}
         \centering
         \includegraphics[width=\textwidth]{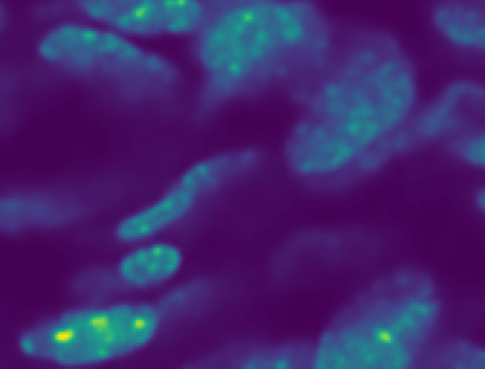}
     \end{subfigure}
     \begin{subfigure}[b]{0.19\textwidth}
         \centering
         \includegraphics[width=\textwidth]{images/Mouse_skull_nuclei_baseline_seed0_0_0_real_B_crop_72_200_32_200.png}
     \end{subfigure}
     
     \par \smallskip
     
     \centering 
     \begin{subfigure}[b]{0.19\textwidth}
         \centering
         \includegraphics[width=\textwidth]{images/Mouse_skull_nuclei_baseline_seed0_0_0_real_A_crop_72_200_32_200.png}
        \caption{Input crop}
     \end{subfigure}
     \begin{subfigure}[b]{0.19\textwidth}
         \centering
         \includegraphics[width=\textwidth]{images/Mouse_skull_nuclei_CLreg0_TV1e-4_SSIM10_seed0_0_0_fake_B_crop_72_200_32_200.png}
         \caption{All training samples}
     \end{subfigure}
     \begin{subfigure}[b]{0.19\textwidth}
         \centering
         \includegraphics[width=\textwidth]{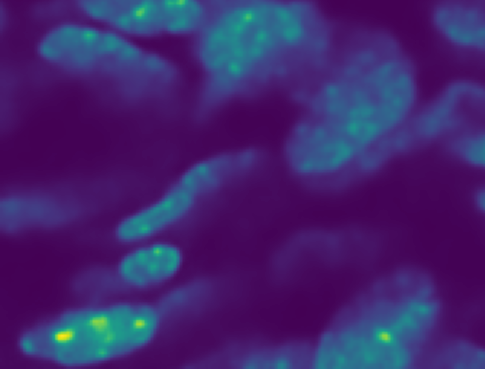}
        \caption{32 training samples}
     \end{subfigure}
     \begin{subfigure}[b]{0.19\textwidth}
         \centering
         \includegraphics[width=\textwidth]{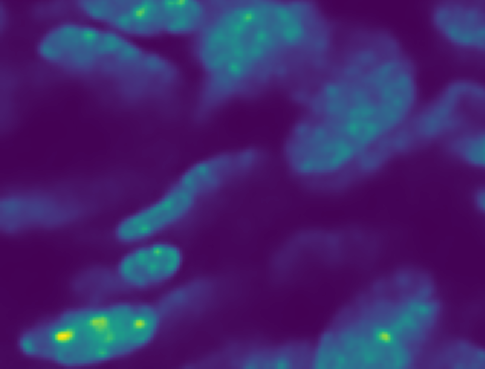}
        \caption{16 training samples}
     \end{subfigure}
     \begin{subfigure}[b]{0.19\textwidth}
         \centering
         \includegraphics[width=\textwidth]{images/Mouse_skull_nuclei_baseline_seed0_0_0_real_B_crop_72_200_32_200.png}
        \caption{Ground truth}
     \end{subfigure}
     
    \caption{\textbf{Generalization study for the mouse \textit{skull nuclei} dataset}. First row shows the baseline configuration of our method, second row the version with contrastive learning, and the third row the complete proposed method. Best viewed on a computer display.}
    \label{fig:qualitative_results_ns_msn}
\end{figure*}

In addition, we perform additional experiments to find out how well our method can generalise with few samples. To test it, we successively reduce the number of traininig examples to 32 and 16 samples, as shown in Table \ref{tab:results_ns}. The original total amount of training examples are 100 for \textit{Convallaria}, 100 for mouse actin and 200 for mouse skull nuclei, as described in Section \ref{ssec:data}. Figures \ref{fig:qualitative_results_ns_conv}, \ref{fig:qualitative_results_ns_ma}, and \ref{fig:qualitative_results_ns_msn} show a qualitative comparison of the denoising method with the three best configuration and varying number of training instances for \textit{Convallaria}, mouse actin and mouse skull nuclei datasets, respectively.

\section{Conclusion}
\label{sec:discussion}

We have introduced a highly competitive generative method for few-shot microscopy image denoising with a great ability to generalize thanks to the use of contrastive learning (see Table \ref{tab:results_sota}).

Our approach combines two well-known methods in image restoration, namely pix2pix and contrastive learning \cite{chen2020simple}. Our experiments demonstrate that this combination leads to significant improvements in the quality of denoised microscopy images even when working with limited amounts of data. In addition, we introduce two loss functions that further enhance the structural appearance of the denoised images, resulting in even better performance. The use of contrastive learning is particularly valuable for scenarios with limited data, where it allows the method to adapt effectively (see Table \ref{tab:results_ns}).

Our ultimate goal is not only to produce high-quality denoised images, but also to enable downstream processing, such as cell counting or segmentation, which is critical for many applications in the biomedical field. Although we focus on per-pixel noise in this work, we believe that our method could potentially handle more complex image degradation models, such as structured noise, blur, or compression artifacts. However, this remains a topic for future research.

Despite its promising results, our method does not improve the state-of-the-art performance on the mouse actin dataset. We hypothesize that this may be due to the high-frequency content of the images, which may not be well-reproduced by the pix2pix architecture. We also acknowledge that newer methods, such as SPADE \cite{park2019semantic}, CC-FPSE \cite{liu2019learning}, and LC-GAN \cite{tang2020local}, outperform pix2pix and may be more suitable for this task. Other methods based in diffusion models or transformers will also be explored in future work, since they have shown remarkable performance in image restoration tasks \cite{wang2022uformer, ali2023vision, zamir2022restormer}.

Finally, while our method requires only a limited amount of paired data, it is still supervised. We aim to make it fully unsupervised in the future or at least remove the need for paired data without diminishing its performance.


%



\section*{Conflict of interest}

The Author(s) declare(s) that there is no conflict of interest.

\section*{Acknowledgment}

This research is supported by the National Research Foundation, Prime Minister’s Office, Singapore under its Campus for Research Excellence and Technological Enterprise (CREATE) programme, France BioImaging infrastructure (ANR-10-INSB-04), the ''programme investissement d’avenir`` (ANR-10-IDEX-03-02) and the CNRS. Computational resources and infrastructure used in present publication were provided by the Bordeaux Bioinformatics Center (CBiB).

\ifCLASSOPTIONcaptionsoff
  \newpage
\fi



%
\bibliographystyle{IEEEtran}
\bibliography{IEEEabrv, mybibfile}

%








\end{document}